\documentclass{emulateapj}

\received{}
\accepted{}

\slugcomment{Accepted by ApJ}
\shortauthors{Kuraszkiewicz et al.}
\shorttitle{PCA of the SED and Emission Line Properties of Red 2MASS AGN}

\def\etal    {{\it et~al.}}

\def\nh      {${\rm N_H}$}
\def\lax     {${_<\atop^{\sim}}$}
\def\gax     {${_>\atop^{\sim}}$}
\def\chandra {{\it Chandra}}
\def\jmk     {{J$-$K$_S$}}
\def\O3      {{[O\,{\small III}]}}
\def\O2      {{[O\,{\small II}]}}
\newcommand{\kms}{\ifmmode~{\rm km~s}^{-1}\else ~km~s$^{-1}~$\fi}

\def\plotfiddle#1#2#3#4#5#6#7{\centering \leavevmode
    \vbox to#2{\rule{0pt}{#2}}
    \includegraphics{#1}}

\begin{document}

\title{PCA of the Spectral Energy Distribution and Emission Line
  Properties of Red 2MASS AGN} 
\author{Joanna Kuraszkiewicz\altaffilmark{1}, Belinda
  J. Wilkes\altaffilmark{1}, Gary Schmidt\altaffilmark{2}, 
 Paul S. Smith\altaffilmark{2}, Roc Cutri\altaffilmark{3}, Bo\.zena
  Czerny\altaffilmark{4}}
\altaffiltext{1}{Harvard-Smithsonian Center for Astrophysics, Cambridge, MA 02138}
\altaffiltext{2}{Steward Observatory, University of Arizona, Tucson, AZ 85721}
\altaffiltext{3}{IPAC, Caltech, MS 100-22, Pasadena, CA 91125}
\altaffiltext{4}{Nicolaus Copernicus Astronomical Center, Warsaw, Poland}

\begin{abstract}

We analyze the spectral energy distributions (SEDs) and emission line
properties of the red (\jmk\ $> 2$) 2MASS AGN using principle
component analysis (PCA). The sample includes 44 low redshift AGN with
low or moderate obscuration (\nh\ $< 10^{23}$~cm$^{-2}$) as indicated
by X-rays and SED modeling. The obscuration of the AGN allows us to
see weaker emission components (host galaxy emission, AGN scattered
light) which are usually outshone by the AGN.  The first four
eigenvectors explain 70\% of the variance in the sample. Eigenvector~1
(33\% variance in the sample) correlates with the ratios of the
intrinsic X-ray flux to the observed optical/IR fluxes and the
F(2$-$10keV)/F([O\,{\small III}]) ratio. We suggest that it is
primarily driven by the $L/L_{Edd}$ ratio and strengthened by
intrinsic absorption (both circumnuclear and galactic).  Eigenvector~2
(18\% of variance) correlates with optical/IR colors (B$-$K$_S$,
B$-$R, \jmk ) and optical spectral type and depends on the
contribution of the host galaxy relative to the observed AGN emission.
Eigenvector~3 (12\% of variance) correlates with reddening indicators
obtained from the X-rays (hardness ratio, spectral index, \nh ) and
the narrow H$\alpha$/H$\beta$ emission line ratio. Their relation
suggests a common absorber for the optical/X-rays lying outside the
narrow-line region (NLR), possibly in a moderately inclined host
galaxy.  Eigenvector~4 (8\% of variance) correlates with the degree of
polarization and the broad H$\alpha$/H$\beta$ ratio, indicating that
dust which scatters the nuclear emission (continuum and the broad-line
region emission) also reddens the broad-lines. Our analysis shows
that, although as suggested by unification schemes, the inclination
dependent obscuration (circumnuclear and the host galaxy) is important
in determining the AGN SEDs, the $L/L_{Edd}$ ratio is the most
important factor, followed by host galaxy emission.

\end{abstract}

\keywords{galaxies: active --- quasars: general}

\section{Introduction}

Principal Component Analysis (PCA) is a mathematical tool used to
reduce the dimensionality of a dataset without losing information. It
transforms the old coordinate system where variables correlate among
each other into a new coordinate system defined by a smaller number of
uncorrelated variables called principal components. The first
principal component accounts for the most variance in the data, and
the successive principal components explain as much of the remaining
variance in the data as possible (see e.g. an introduction to PCA by
Francis \& Wills 1999 or Murtagh \& Heck 1987).

Since QSOs show a plethora of emission line and continuum properties, 
PCA is ideal for studying their properties and understanding which are
most important. PCA was first introduced to the AGN world by Boroson
\& Green (1992; hereafter BG92) who analyzed the optical emission line
and continuum properties of a complete, low redshift, Bright Quasar
Survey (BQS) sample.  It was found that the primary eigenvector (EV1)
is anticorrelated with Fe\,{\small II}\,$\lambda 4570$ strength
(equivalent width and Fe\,{\small II}/H$\beta$ ratio), correlated with
[O\,{\small III}]\,$\lambda5007$ strength (luminosity and peak) and
H$\beta$ FWHM, and anticorrelated with the blue asymmetry of the
H$\beta$ line. Hence BG92 concluded that EV1 is driven by a physical
parameter that ``... balances Fe\,{\small II} excitation against
illumination of the narrow line region''.  The BG92 EV1 was also found
to correlate with the soft X-ray properties: luminosity and spectral
index (Boroson \& Green 1992; Corbin 1993; Laor et al. 1994, 1997,
Wang \etal\ 1996, Brandt \& Boller 1998) and the correlations were
found to be stronger than with the individual emission line parameters
(Brandt \& Boller 1998). This implied that EV1 has a more fundamental
physical meaning.  Physical parameters that have been suggested to
drive EV1 include: the $L/L_{Edd}$ ratio, orientation, and black hole
spin.  Since EV1 is correlated with [O\,{\small III}] emission,
thought to be an isotropic property, BG92 concluded that EV1 is not
driven by orientation. Although the isotropy of the [O\,{\small III}]
emission has since been called into question (Jackson \& Browne 1990,
Di Serego Alighieri et al. 1997, Tadhunter et al. 1998) the
correlation of EV1 with the inclination independent [O\,{\small
II}]\,$\lambda3727$ emission, originating at larger distances from the
nucleus than [O\,{\small III}], strongly supports the EV1 orientation
independence (Kuraszkiewicz \etal\ 2000), at least for the radio-quiet
subsample of the BQS.  BG92 suggest that the $L/L_{Edd}$ ratio is the
primary driver of EV1.

BG92 EV1 was also found to correlate with the UV emission line
properties: CIII] width, SiIII]/CIII] ratio, CIV and NV strength, CIV
line shift (Wills \etal\ 1999, Wills, Shang, \& Yuan 2000, Sulentic
\etal\ 2000), suggesting that EV1 is also linked to the physics of
fueling and outflow and that higher accretion rates are linked to
denser gas and possibly higher nuclear starburst activity (Wang
\etal\ 2006).

By combining observational data together with numerical simulations,
Marziani \etal\ (2001) and Sulentic \etal\ (2000) were able to
reproduce the correlations between optical emission line properties
found in EV1 and confirm that EV1 is related to $L/L_{Edd}$, but
suggest that, in their sample, orientation is also important. Their
sample includes both the BQS quasars from BG92 (face-on, low
obscuration), and inclined/reddened sources of intermediate type and
Type~2 AGN that have {\it FOS/HST} spectra.

A different way to approach the analysis of AGN properties is spectral
PCA first applied by Francis et al. (1992) to the analysis of more
than 200 spectra from the Large Bright QSO Survey (LBQS; redshift
range: $0.2\le z < 3.4$)   
or by Shang \etal\ (2003) to the spectra of PG QSOs. Their analysis is
superseded by Yip \etal\ (2004) who analyze spectra of more than
16,000 QSOs from the Sloan Digital Sky Survey (SDSS) with a wider
range of redshifts ($0<z<5.4$), inclinations and/or reddening
values. Their spectral PCA found the following principal components:
mean spectrum, host galaxy component, UV--optical continuum slope, and
correlations of Balmer emission lines.  Although the results obtained
from spectral PCA and emission line+SED PCA are difficult to compare,
it is interesting to invoke them here for comparison with our results.

In this paper we analyze a sample of red (\jmk\ $>2$) low-redshift
2MASS AGN with moderate absorption \nh \lax $10^{23}$cm$^{-2}$. Their
SEDs were analyzed in detail in Kuraszkiewicz \etal\ (2008; hereafter
Paper~I) and the emission line properties are presented in
Kuraszkiewicz \etal\ (2008 in preparation; hereafter Paper~II).  In
this paper we run PCA on the SED and optical emission line correlation
matrix and present our interpretation of the eigenvectors in this
moderately absorbed 2MASS AGN sample. It is important to state here
that the results of the PCA analysis are dependent on the selection of
the sample and chosen properties. We do not expect the same
eigenvectors in the analysis of the 2MASS AGN sample as those found by
BG92 in the unobscured, face-on PG QSO sample. The PG QSOs' first two
eigenvectors were dominated by the parameters of the central engine
($L/L_{Edd}$ and accretion rate -- Boroson 2002), while our 2MASS AGN
sample, due to AGN obscuration, is well suited for studies of weaker
components such as the host galaxy emission and intrinsic absorption.

\section{The Sample and its Properties}

The sample includes 44 red (\jmk\ $>2$) 2MASS AGN selected to have
B$-$K$_S>4.3$, and $K_S<13.8$, and observed by \chandra\ (see Paper~I
for more details). The sources have K band luminosities comparable to
those of quasars (Cutri \etal\ 2002) and lie at low redshift ($z <
0.37$). The sample shows a wide range of observed
$K_S$-band-to-X-ray(1keV) slopes (1.1\lax $\alpha_{KX}$\lax 2; Wilkes
\etal\ 2002), and a broad range of observed polarization fraction at R
band ($0\% < P < 13\%$; Smith \etal\ 2002, 2003).  As found in Paper~I
the sample's median SED is redder than that of the blue
optically/radio selected QSOs (Elvis \etal\ 1994) or the hard--X--ray
selected AGN (Kuraszkiewicz \etal\ 2003), showing little or no big
blue bump and implying that near-IR color selection isolates the
reddest subset of AGN that can be classified optically. Modeling of
the near-IR/optical colors and X-ray spectral analysis in Paper~I
implies moderate amounts of reddening (\nh\ $< few \times
10^{22}$~cm$^{-2}$). The more highly obscured AGN in this sample are
most likely viewed at intermediate angles, where the central engine is
obscured from our view by either circumnuclear dust or dust lying
further out in the host galaxy inclined to our line of sight. The
obscuration allows us to see/study the contributions of weaker
emission components such as host galaxy emission or AGN scattered
light emission which are normally outshone by the bright AGN. The
sample includes: 7 Type~1 AGN with optical spectra resembling NLS1 or
BALQSOs, 11 Type~2s with low (for a Type~2) \nh\ $\sim
10^{22}$cm$^{-2}$, and 26 intermediate type sources. Many of the 2MASS
AGN studied here also show high optical polarization ($P > 4$\%).

\section{Principle Component Analysis Results}

In order to investigate the relative importance of various parameters
in determining the SED shapes and emission--line properties in the red
2MASS sources, we analyze the relations between the continuum
parameters and emission-line parameters by running principal component
analysis (PCA) on the correlation matrix.  The first PCA run included
all continuum properties analyzed in Paper~I (Sections~4,5,6): optical
and near-IR colors (B$-$R, B$-$K$_S$, \jmk ), optical type, optical
slope, $\alpha_{opt}$, from optical spectra, degree of polarization,
redshift, X-ray slope, intrinsic absorption \nh , X-ray hardness
ratio, X-ray (1keV) luminosity (all from \chandra\ spectral fitting),
bolometric luminosity from SEDs, various IR/optical/X-ray flux or
luminosity ratios\footnote{Flux/luminosity ratios included:
F(1keV)/F$_B$, F(1keV)/F$_R$, F(1keV)/F$_I$, F(1keV)/F$_K$,
L(0.2-0.4$\mu$m)/L(0.4-0.8$\mu$m), L(0.2-0.4$\mu$m)/L(0.8-1.6$\mu$m),
L(0.4-0.8$\mu$m)/L(0.8-1.6$\mu$m), L(1-10$\mu$m)/L(10-100$\mu$m),
L(3-60)$\mu$m)/L(60-100$\mu$m), L(0.8-1.6$\mu$m)/L(1-100$\mu$m),
L(0.8-1.6$\mu$m)/L(60-100$\mu$m), L(0.8-1.6$\mu$m)/L(3-60$\mu$m)}, and
the emission--line parameters from Paper~II: $W_{\lambda}$, FWHM,
shift from systemic redshift of: H$\beta$ (broad and narrow
components), H$\alpha$ (broad and narrow components), [O\,{\small
III}], [O\,{\small II}], optical Fe\,{\small II}, and the following
line ratios: [O\,{\small II}]/[O\,{\small III}],
H$\beta^{broad}$/[O\,{\small III}] , H$\beta^{narrow}$/[O\,{\small
III}], H$\alpha^{narrow}$/H$\beta^{narrow}$,
H$\alpha^{broad}$/H$\beta^{broad}$, Fe\,{\small II}/H$\beta^{broad}$,
Fe\,{\small II}/H$\beta^{narrow}$, Fe\,{\small II}/[O\,{\small
III}]. A total of 56 parameters were analyzed.  The EW, FWHM, line
shifts, most line ratios, and optical--to--IR and IR--to--IR
luminosity ratios did not contribute to the first few eigenvectors and
running PCA with or without them gave similar results (i.e. nearly
identical eigenvectors). To simplify the analysis these parameters
were omitted from the final PCA run. The final set of parameters is
shown in the first column of Table~1. Columns 2--5 show
their relation with the first four eigenvectors (bold font denoting
parameters that dominate the eigenvector) and quote in percentages how
much variance they explain. Together the first four eigenvectors
explain 70.3\% of the variance in the sample. Their interpretation is
presented below.

\subsection {Eigenvector~1}

\begin{figure}
\epsscale{0.8}
\plotone{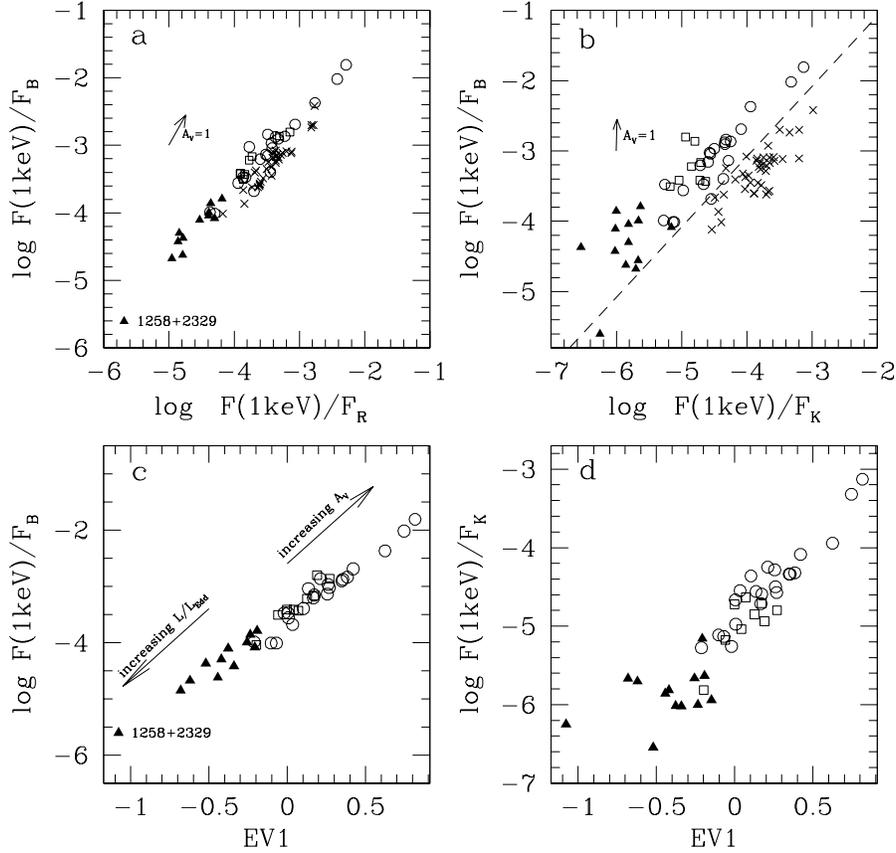}
\caption{{\it a, b} $-$ Correlations between the intrinsic 1~keV X-ray
 flux to observed B flux ratio and the intrinsic 1~keV X-ray flux to
 observed R and K flux ratios. These are examples of correlations
 found to compose EV1.  The X-ray flux is corrected for Galactic and
 intrinsic extinction while the B,R and K fluxes are corrected only
 for Galactic extinction. Crosses represent the blue optically/radio
 selected quasars from Elvis \etal\ (1994). The 2MASS objects in all
 figures are delineated by their Chandra S/N classifications, with
 circles, squares and filled triangles representing sources with {\it
 C} (highest S/N), {\it B} (medium S/N) and {\it A} (low S/N) spectral
 types respectively.  The dashed line in Fig.~1b shows the
 B$-$K$_S$=4.3 color cut used to select the red 2MASS AGN. 1258+2329
 lies below this cut since SuperCOSMOS B magnitude (which is more
 consistent with the SED) is used here instead of the USNO--A2
 magnitude originally used for color selection. {\it c, d} $-$
 Correlations between Eigenvector~1 and the ratio of intrinsic 1keV
 X-ray flux to the observed flux in the B and K bands.}
\label{fig:Fx/FBvsFx/Fnn}
\end{figure}

\begin{figure}
\plotone{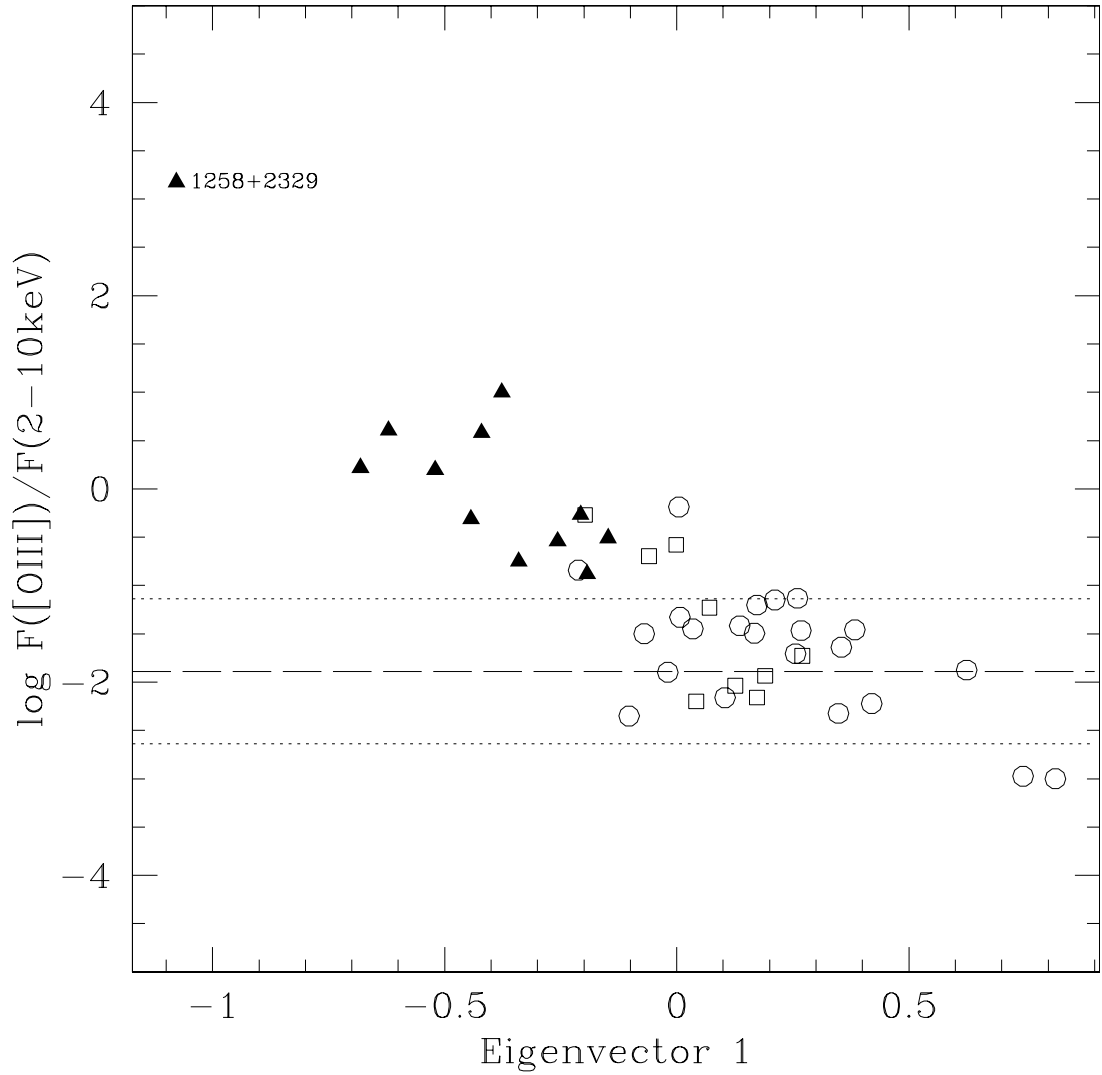}
\caption{Correlation between the F([O\,{\small III}])/F(2$-$10keV)
  ratio and Eigenvector~1.  X-ray fluxes are corrected for Galactic
  and intrinsic reddening. The [O\,{\small III}] emission--line flux
  is corrected for NLR reddening estimated using the narrow
  $H\alpha/H\beta$ ratio. Symbols indicate X-ray S/N as in
  Fig.~\ref{fig:Fx/FBvsFx/Fnn}.  Dashed line indicates the value of
  $\log F([O\,III])/F(2-10keV) = -$1.89$\pm$0.75 (3$\sigma$
  uncertainty - dotted line) found by Mulchaey \etal\ (1994) for
  Seyfert~1s and Compton-thin Seyfert~2s.}
\label{fig:EV1}
\end{figure}

Most of the variance (32.9\%) in the sample is due to the correlations
between the ratios of the intrinsic 1~keV X-ray flux to the observed
optical/IR flux: $F(1keV)/F_B$, $F(1keV)/F_R$, $F(1keV)/F_I$,
$F(1keV)/F_J$, $F(1keV)/F_K$ (see Fig.~\ref{fig:Fx/FBvsFx/Fnn}).  We
choose to use the observed optical and IR fluxes since, in this
moderately obscured/inclined sample, these fluxes depend on intrinsic
reddening and host galaxy emission, which are a priori unknown and
hence difficult to subtract. The X-ray fluxes, on the other hand, are
corrected for intrinsic absorption, which was estimated from the
\chandra\ X-ray spectral fitting in the following way: the higher
(\gax 80) count sources were fit with a power-law and rest-frame
absorption with both parameters free (hereafter referred to as the
``{\bf C}'' fits), the medium ($\sim 30-80$) count sources, were fit
with a fixed power-law with an X-ray photon index $\Gamma=2$ and \nh\
free (``{\bf B}'' fits).  For the lowest count sources (\lax 30
counts; ``{\bf A}'' fits), fits were made using a power-law with fixed
$\Gamma = 2$ and \nh\ $= 7.6\times10^{21}$~cm$^{-2}$ (the median
absorption from the {\it C} fits). The importance of the optical/X-ray
flux ratio can be understood, for example, in the disk-wind model
paradigm (Murray \& Chiang 1990, Proga \etal\ 2005, and references
therein), which predict coupling between processes producing X-ray and
optical/UV photons.  EV~1 is also strongly anticorrelated with the
$F([O\,{\small III}])/F(2-10keV)$ ratio (see Fig.~\ref{fig:EV1}).

\begin{figure}
\plotone{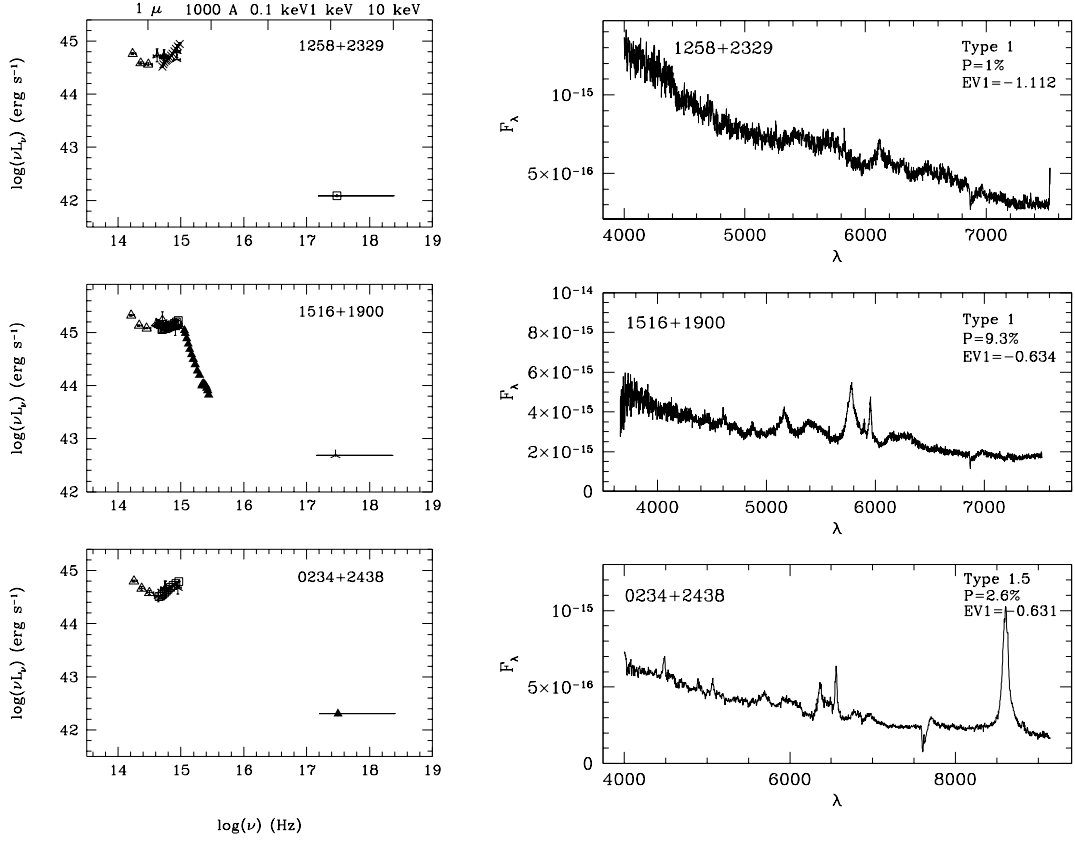}
\caption{SED and emission line properties of AGN with the lowest EV1
  values. {\it Left column} $-$ Restframe far--IR to X-ray SEDs. X-ray
  fluxes are corrected for Galactic and intrinsic absorption. IR-to-UV
  wavelengths have not been corrected for reddening or host galaxy
  contamination. SEDs show low F(1keV)/F$_{B,R,J,K}$ ratios and blue
  B$-$R colors. {\it Right column} $-$ Observed-frame optical spectra
  on $F_{\lambda}$ (erg~s$^{-1}$~cm$^{-2}$~\AA$^{-1}$) versus
  $\lambda$ (\AA ) scale. Optical spectral type, degree of optical
  polarization and EV1 value have been denoted on each spectrum.
  Spectra resemble those of NLS1s and BALQSOs characterized by strong
  Fe\,{\small II} and weak [O\,{\small III}] emission.}
\label{fig:extEV1low}
\end{figure}

\begin{figure}
\plotone{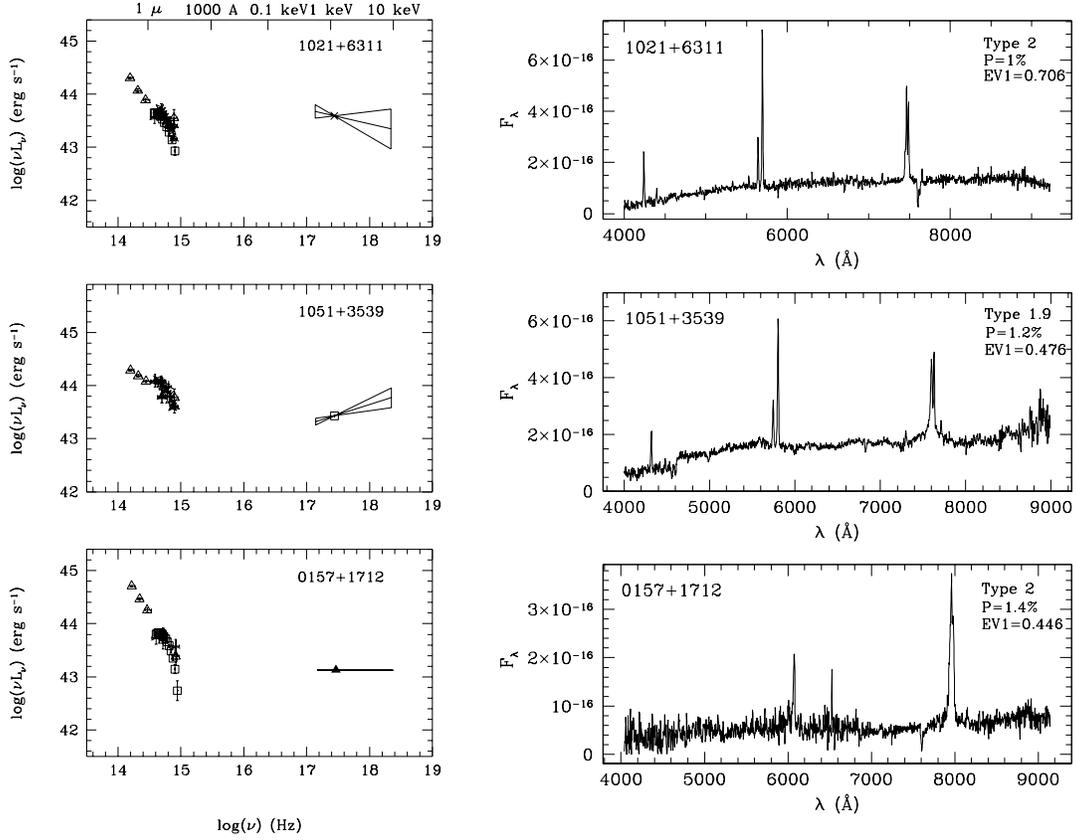}
\caption{SED and emission line properties of AGN with the highest EV1
  values. {\it Left column} $-$ Restframe far--IR to X-ray SEDs (for
  details see Fig.~\ref{fig:extEV1low}). SEDs show high
  F(1keV)/F$_{B,R,J,K}$ ratios. {\it Right column} $-$ Observed-frame
  optical spectra on $F_{\lambda}$ (erg~s$^{-1}$~cm$^{-2}$~\AA$^{-1}$)
  versus $\lambda$ (\AA ) scale. Optical spectral type, degree of
  optical polarization and EV1 value have been denoted on each
  spectrum. Spectra are characterized by strong narrow lines and weak
  Balmer and Fe\,{\small II} lines.}
\label{fig:extEV1high}
\end{figure}

To understand the meaning of EV1 we need to take a closer look at the
objects with extreme values of EV1. Their SEDs and optical spectra are
presented in Figures~\ref{fig:extEV1low} and \ref{fig:extEV1high}, and
their properties are summarized in Table~2. We find that the most
negative EV1 sources (Fig.~\ref{fig:extEV1low}) with lowest
$F(1keV)/F_{B,R,I,J,K}$ and highest F([O\,{\small III}])/F(2$-$10keV),
are Type~1--1.5 with blue optical spectra dominated by strong
Fe\,{\small II} and weak (relatively to H$\beta$) [O\,{\small III}]
emission, that resemble the spectra of NLS1s and BALQSOs. Although
their low S/N X-ray spectra ({\it A} fits) do not yield an estimate
for \nh , modeling of B$-$R and \jmk\ colors in Paper~I (Section~5)
gave low \nh\ \lax $5\times10^{21}$~cm$^{-2}$ (hence their X-ray
fluxes are likely overestimated since \nh\ $=
7.6\times10^{21}$~cm$^{-2}$ was assumed for these {\it A} fits during
spectral fitting) and shows that the SEDs at optical wavelengths are
dominated by AGN emission ($>$94\% of total observed flux at R band is
due to the AGN in these sources).

The most positive EV1 sources (Fig.~\ref{fig:extEV1high}) with highest
$F(1keV)/F_{B,R,I,J,K}$ and lowest F([O\,{\small III}])/F(2$-$10keV),
are Type~1.9--2, with red spectra showing strong [O\,{\small III}] and
[O\,{\small II}] emission, and weak Fe\,{\small II} and H$\beta$
emission and strong galactic features.  X-ray spectral fitting gives
\nh\ $\sim 10^{22}$~cm$^{-2}$ consistent with reddening obtained from
the modeling of B$-$R and \jmk\ colors in Paper~I (Section~5), that
also shows high host galaxy contribution at optical wavelengths
($>$61\% of the total observed flux at R band is due to host galaxy).

The luminosity of the [O\,{\small III}]\,$\lambda$5007 emission line,
originating from the narrow-line region (NLR), has been suggested as
an indicator of the intrinsic nuclear luminosity of the AGN due to the
similarity of the [O\,{\small III}]-to-hard-X-ray flux ratio between
Seyfert~1 and Seyfert~2 galaxies (Mulchaey \etal\ 1994, Alonso-Herrero
\etal\ 1997, Turner \etal\ 1997).  As presented in Fig.~\ref{fig:EV1}
the most negative EV1 sources do not follow the [O\,{\small III}] to
hard-X-ray relation showing higher F([O\,{\small III}])/F(2$-$10keV)
ratios. This indicates that in these sources the intrinsic X-rays are
weak. The most positive EV1 sources, on the other hand, have lower
F([O\,{\small III}])/F(2$-$10keV) indicating strong intrinsic X-rays.

\begin{figure}
\plotone{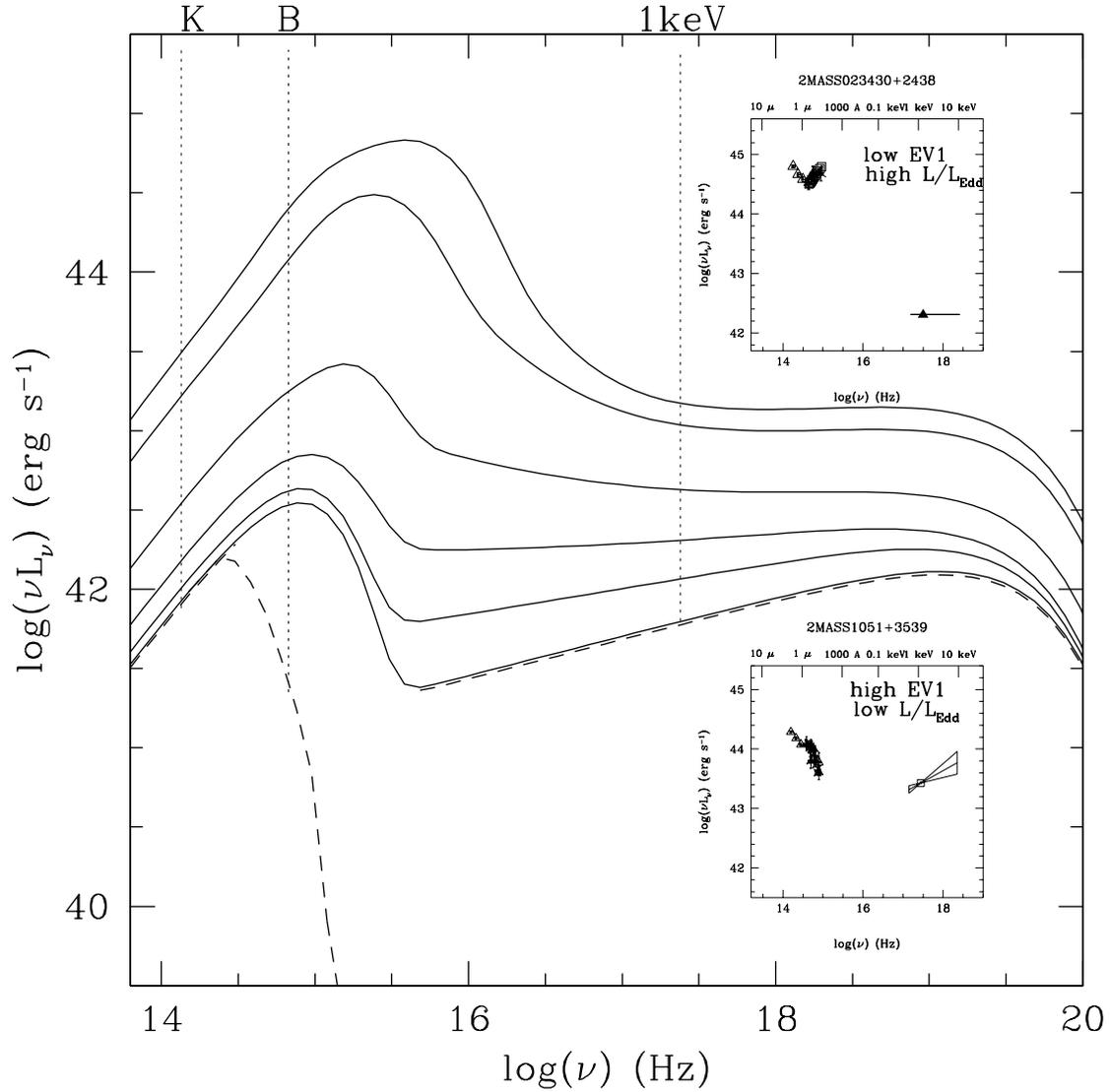}
\caption{The AGN spectral energy distribution predicted by the
  accretion disk and corona model of Witt, Czerny \& \.Zycki (1996)
  for a black hole mass 10$^8$~M$_{\odot}$, viscosity parameter
  $\alpha=0.33$ and from top to bottom $L/L_{Edd}$=0.25, 0.1, 0.01,
  0.003, 0.0017, 0.0013 respectively. Higher $L/L_{Edd}$ give a more
  pronounced big blue bump relative to the X-rays and lower
  F(1keV)/F$_{B,R}$ ratios. Lower $L/L_{Edd}$ produce a weaker big
  blue bump relative to the X-rays and higher F(1keV)/F$_{B,R}$
  ratios.  The dashed line shows the $L/L_{Edd}$=0.0013 model reddened
  at optical/UV wavelengths by A$_V$=3 (a reddening value obtained
  from IR/optical color modeling for 1051+3539, shown in lower insert;
  X-ray flux is left intrinsic since the \chandra\ X-ray flux was
  corrected for both Galactic and intrinsic absorption).  Inserts show
  examples of red 2MASS AGN SEDs with high and low Eigenvector~1 values.}
\label{fig:models}
\end{figure}

The differences in X-ray output relative to optical/UV may be
explained for example by an accretion disk and corona model of Witt,
Czerny, \& \.Zycki (1997). In this model both the accretion disk and
corona accrete and generate energy through viscosity, and division of
the flow into optically thin (corona) and optically thick (disk)
regions results from the cooling instability discussed by Krolik,
McKee and Tarter (1981).
The model is defined by 3 parameters: the mass of the central black
hole, the ratio of the luminosity to the Eddington luminosity
$L/L_{Edd}$ (or the accretion rate \.M), and the viscosity parameter
($\alpha_{vis}$) assumed to be the same in both the disk and the
corona.  The model predicts a systematic change in the opt/UV/X-ray
SED as a function of $L/L_{Edd}$.  For high $L/L_{Edd}$, the big blue
bump is stronger both in luminosity and relative to the X-rays
so the $F(1keV)/F_B$ ratio is low (as in low EV1 sources). For low
$L/L_{Edd}$ the big blue bump is weaker and the $F(1keV)/F_B$ ratio
becomes high (as in high EV1 sources; see Fig.~\ref{fig:models}).  
Since the accretion disk and corona model predictions match those with
EV1, they suggest that EV1 is fundamentally related
(i.e. anti-correlated) to the $L/L_{Edd}$ ratio.  This is further
strengthened by the fact that the most negative EV1 sources (with
lowest $F(1keV)/F_B$) show optical spectra that characterize NLS1s and
BALQSOs (strong Fe\,{\small II} and weak [O\,{\small III}] emission)
that are thought to have high $L/L_{Edd}$ (e.g. Pounds, Done \&
Osborne 1995, Boller, Brandt, Fink 1996, Kuraszkiewicz \etal\ 2000).

From Table~1 and Fig.~\ref{fig:Fx/FBvsFx/Fnn}c,d we see that the
correlation between EV1 and the F(1keV)/F$_{B}$ ratio is stronger than
the correlation between EV1 and F(1keV)/F$_{K}$.  The optical/IR
fluxes used to calculate these flux ratios are observed, not
intrinsic, so they are affected by dust reddening (and to a lesser
extent also by host galaxy emission). Fluxes at optical wavelengths
will be more affected than those at near-IR wavelengths. Since EV1
shows stronger correlation with the $F(1keV)/F_B$ ratio, which depends
both on the $L/L_{Edd}$ and reddening, and shows more scatter with the
$F(1keV)/F_K$ ratio, which depends mostly on the $L/L_{Edd}$, and less
on reddening, we conclude that EV1 is dominated by
$L/L_{Edd}$. However, dust obscuration also contributes as it
stretches the underlying $L/L_{Edd}$ driven correlation, at optical
wavelengths, in these moderately obscured 2MASS sources.  If dust
obscuration is inclination dependent then this result is consistent
with Marziani \etal\ (2001) who suggest that their EV1 found for a
heterogeneous sample of mixed type (1, 2 and intermediate) AGN, is
related to the $L/L_{Edd}$ ratio convolved with orientation.


\subsection{Eigenvector~2}

\begin{figure}
\plotone{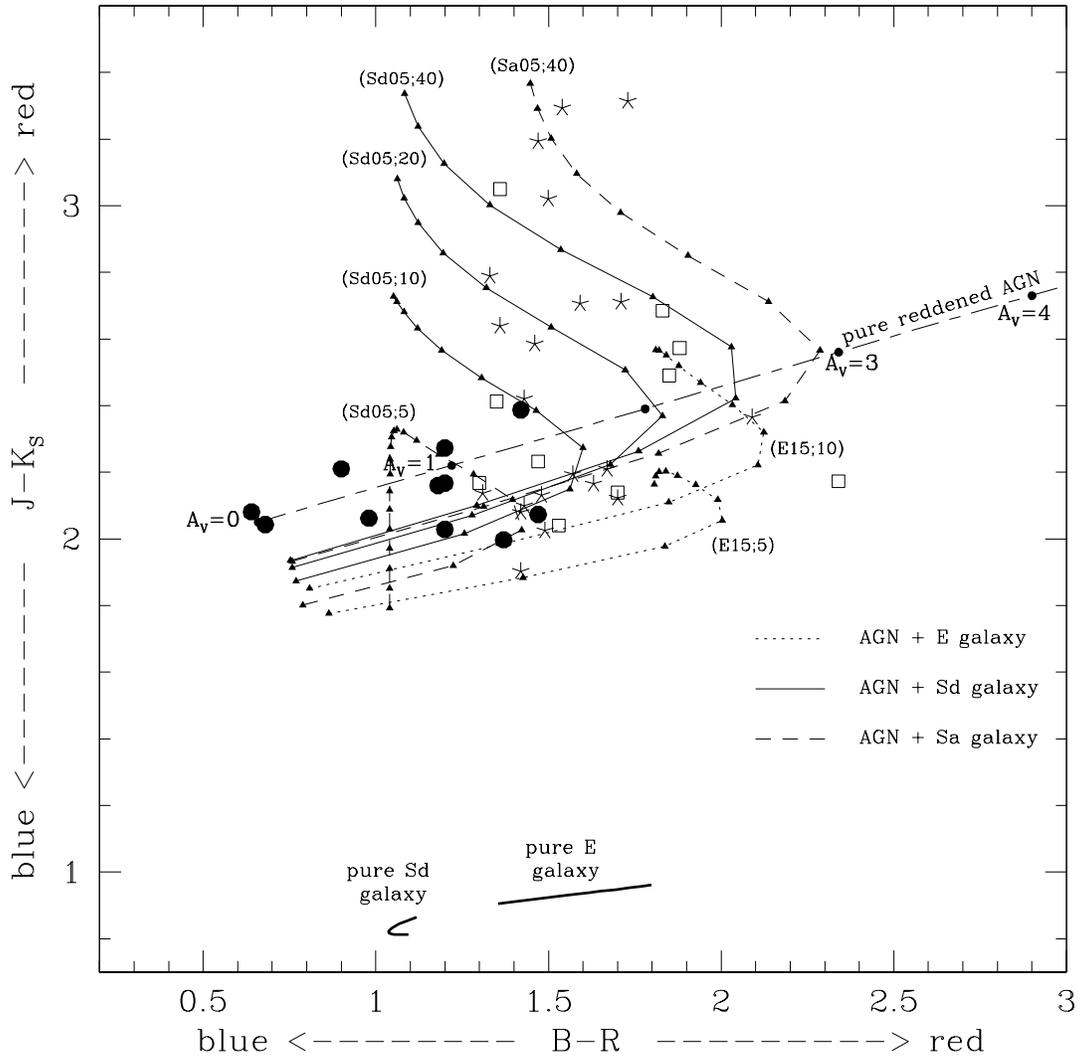}  
\caption{Colors of a reddened AGN combined with contributions from a
  host galaxy. Long-dash-short-dash curve shows colors of a pure
  reddened AGN (Elvis \etal\ 1994 median QSO SED reddened by Milky Way
  dust). Dotted (green) lines represent AGN colors modified by
  emission from an elliptical host with a 15--Gyr stellar population
  (E15; reddest in B$-$R).  Solid (blue) lines represent an AGN~+~Sd
  host and 5--Gyr stellar population (Sd05; bluest in
  B$-$R). Short-dashed (red) line: AGN~+~Sa host galaxy (B$-$R colors
  intermediate between E15 and Sd05).  Each curve starts at the colors
  of an unreddened (A$_V$=0) AGN~+~host galaxy (bluest optical/IR
  colors) and extends to the reddened (A$_V$=10) AGN~+~host galaxy in
  steps of 1~mag.  denoted by small triangles on the curves (curve
  (Sd05;5) has been extended to A$_V$=20). Numbers in parenthesis give
  the host galaxy type and age of stars followed by the intrinsic,
  unreddened AGN/host galaxy flux ratio at R band (e.g. (E15;10):
  elliptical host, 15--Gyr stars, intrinsic AGN/host = 10). Thick
  solid curves indicate pure host galaxy (E and Sd) colors changing
  with star age (from 1 to 15--Gyr). Overplotted are the observed
  colors of the red 2MASS AGN, divided according to the amount of host
  galaxy contribution at R band obtained from color-color modeling:
  where $\le 20$\% (filled circles), 20-60\% (squares), and $>60$\%
  (stars) of the total observed flux at R band is due to host galaxy
  emission.}
\label{fig:colors}
\end{figure}

Eigenvector~2 (hereafter EV2), which explains 17.8\% of the variance
in the sample, is dominated by the observed optical/IR colors (in
order of importance): B$-$K$_S$, B$-$R, \jmk , and optical spectral
type (see Table~1).  In Paper~I we found that the optical B$-$R color,
and to a lesser extent the near-IR \jmk\ color in the 2MASS AGN
sample, are affected by reddening and host galaxy emission (and in a
few, highly polarized objects, also by scattered AGN light). We recall
the results of the optical/IR color modeling from Paper~I in
Fig.~\ref{fig:colors}. The colors of a pure reddened AGN (median QSO
SED from Elvis \etal\ 1994 reddened by Milky Way dust;
long-dash--short-dash line) are modified by host galaxy emission
(Buzzoni 2005 templates), where
a range of host galaxy strengths relative to the intrinsic, {\it
unreddened} AGN at R band was assumed (intrinsic AGN/host galaxy ratio
$= 5-40$). For $A_V$\lax 3 the optical/near--IR colors are dominated by
the AGN. At higher $A_V$ the B$-$R colors become bluer and more
consistent with the host galaxy colors (represented by thick solid
lines).  The \jmk\ colors are much less dependent on reddening and
host galaxy, only visible if their contribution is strong i.e. A$_V >
10$ and AGN/host galaxy ratio low (compare for example (Sd05;5) and
(Sd05;40) curves that have a high and low host galaxy contribution
respectively).

In Table~2 we summarize the properties of extreme EV~2 objects. We
find that the most positive EV2 objects are the bluest in B$-$R,
implying a strong big blue bump. They also have the bluest B$-$K$_S$
colors in sample. They are Type~1s, with optical spectra showing
strong Fe\,{\small II} and weak [O\,{\small III}] emission, resembling
spectra of NLS1 and BALQSO. Column densities derived from X-ray
spectral fitting are low (\lax 1.5$\times 10^{21}$~cm$^{-2}$), and the
optical/IR colors are modeled with a pure, unreddened AGN with no host
galaxy contribution.

The most negative EV2 objects are red in B$-$K$_S$, B$-$R and \jmk\
color. The red optical/IR colors are modeled with an AGN reddened by
dust with $A_V$=10--11~mag. Such reddening is sufficient to fully
obscure the AGN at optical wavelengths, giving way to a large (98\% at
R band) contribution from the host galaxy. At near-IR wavelengths the
reddened AGN colors still dominate in these sources. The red \jmk\ $>
3$ color can only be produced by a reddened AGN and not the host
galaxy which has bluer colors: \jmk\ = 0.8$-$1 (see
Fig.~\ref{fig:colors}).

\begin{figure}
\plotone{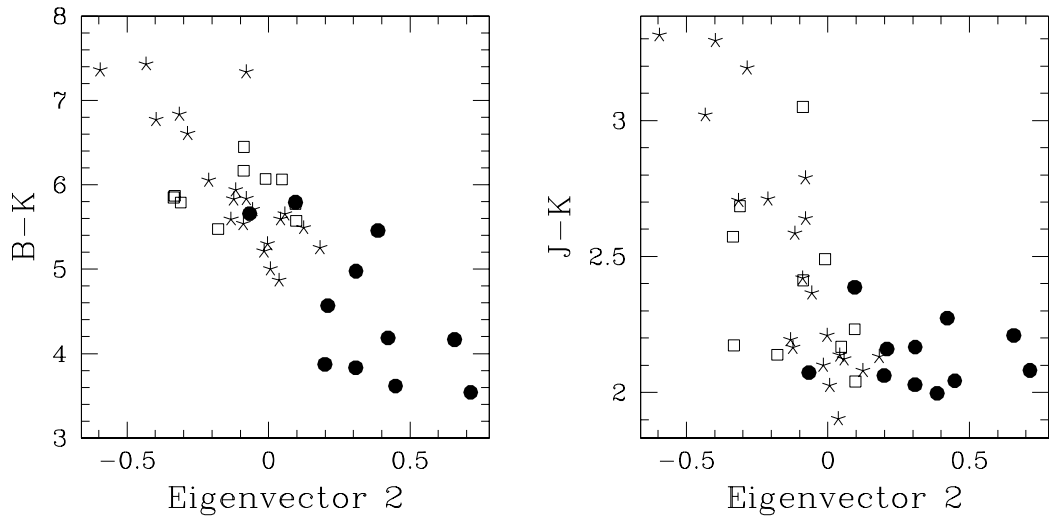}
\caption{Correlations between Eigenvector~2 and the B$-$K$_S$ and
  \jmk\ colors. Symbols as in Fig.~\ref{fig:colors} denote host galaxy
  contribution.}
\label{fig:EV2}
\end{figure}

In Fig.~\ref{fig:EV2} we divide the sources according to the amount of
host galaxy contribution relative to the total emission at R band
obtained from the optical/IR color modeling (numbers are from Table~7
in Paper~I).  We find that the amount of host galaxy contribution
relative to the reddened AGN decreases with increasing EV2
values. This implies that EV2 depends on the amount of host galaxy
contribution, relative to the observed/reddened AGN, where at one
(negative) end of EV2 reside pure, unreddened AGN dominating the
output at optical/near-IR wavelengths, and on the other (positive) end
reside reddened AGN whose output at optical wavelengths is dominated
by host galaxy emission. A stronger correlation of EV2 with the B$-$R
color than the \jmk\ color demonstrates that the B$-$R color is much
more sensitive to host galaxy contamination than the \jmk\ color, as
shown by the color-color modeling (see Fig.~\ref{fig:colors} or
detailed discussion in Paper~I).  Yip \etal\ (2004), who analyze the
SDSS QSO spectra (sample with a wide range of redshifts, inclination
angles and reddening values) using {\it spectral} PCA, also find that
their second principle component is related to the host galaxy.

\subsection{Eigenvector~3}

\begin{figure}
\plotone{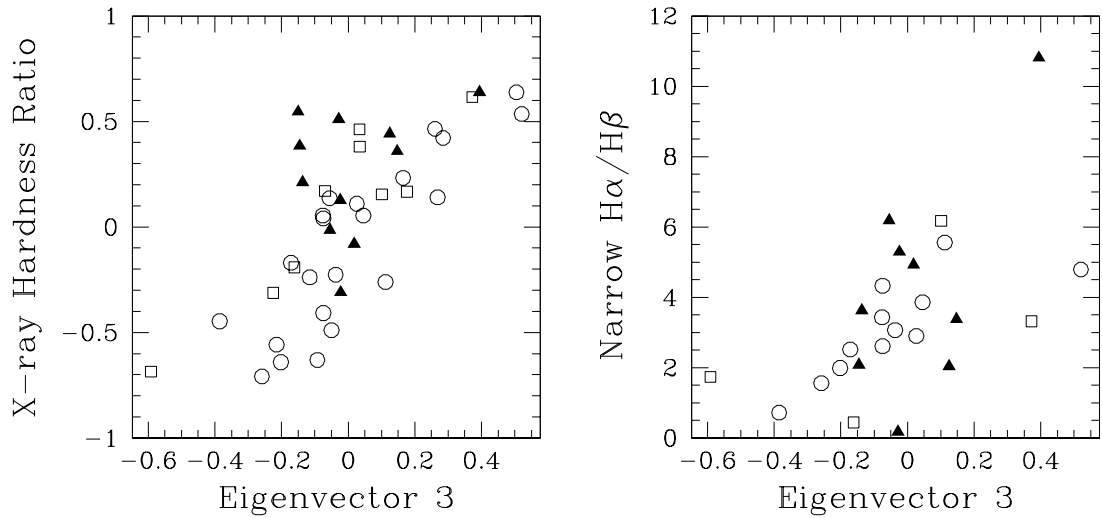}
\caption{Correlations between Eigenvector~3 and the X-ray hardness
  ratio and between Eigenvector~3 and H$\alpha$/H$\beta$ measured from
  narrow emission lines. Symbols indicate X-ray S/N, as in Fig.~\ref{fig:Fx/FBvsFx/Fnn}.}
\label{fig:EV3}
\end{figure}

Eigenvector~3 (hereafter EV3; 11.6\% of the variance) is dominated (in
order of importance) by: X-ray hardness ratio, X-ray spectral index
$\Gamma_X$, column density \nh\ (measured only in the high and medium
S/N \chandra\ spectra i.e. in the {\it C} and {\it B} fits) and the
narrow H$\alpha$/H$\beta$ emission line ratio.  In Fig.~\ref{fig:EV3}
we show two representative correlations between EV3 and the X-ray
hardness ratio and the narrow H$\alpha$/H$\beta$ emission line ratio.
All of these parameters are related to the amount of obscuration in
X-rays or optical. Their relation implies a common absorber for the
optical and X-rays, lying outside the NLR, possibly in a host galaxy
inclined to our line of sight (as in Polletta \etal\ 2008, Deo \etal\
2007). This finding is supported by the HST/WFPC2 I-band images of the
2MASS AGN, which show that these sources reside mostly in host
galaxies with inclination angles of $i=50-75^{o}$ (mean i=67$^{o}$;
Marble \etal\ 2003).  We conclude that EV~3 is likely dominated by
host galaxy absorption.

The presence of redshift as a component of EV3 (Table~1) is secondary
to the presence of \nh , for which measurements become less sensitive
as redshift increases and the absorption is shifted out of the soft
X-ray band. The correlation coefficient reduces significantly if \nh\
is corrected for redshift (assuming it is intrinsic).


\subsection{Eigenvector~4}

Eigenvector~4 (hereafter EV4; 8.0\% of the variance) is dominated by
the degree of optical polarization, broad H$\alpha$/H$\beta$ emission
line ratio and redshift. The relation between the polarization and the
broad H$\alpha$/H$\beta$ ratio implies that the same dust scatters the
intrinsic continuum+broad lines {\it and} reddens the broad emission
lines. Since there is no dependence on the narrow H$\alpha$/H$\beta$
ratio in this eigenvector, this dust must lie outside the BLR but not
outside the NLR, i.e., either on the BLR/NLR border, or within the
NLR.  This is consistent with the conclusions reached by Smith \etal\
(2003) based on spectropolarimetry of the 2MASS AGN.  The correlation
between EV4 and redshift is due to a correlation between redshift and
polarization (chance probability $\sim$ 2\%; there is no correlation
with the broad H$\alpha$/H$\beta$ ratio - chance probability
$>$60\%). This may be a selection effect as higher redshift sources
will have a higher measured degree of polarization due to the
wavelength dependence of dust scattering and lower host galaxy
dilution as one moves into restframe UV where the stellar contribution
decreases.

\section{Conclusions}

PCA analysis of the SED and emission line parameters of a sample of 44
red \jmk\ $> 2$, moderately obscured (\nh\ $< 10^{23}$~cm$^{-2}$)
2MASS AGN shows that 70\% of the variance in the sample is explained
by four eigenvectors, each having a physical explanation:

\begin{itemize}
\item{Eigenvector~1 (33\% of variance in sample) correlates with the
intrinsic X-ray to observed optical, IR and [O\,{\small III}] flux
ratios, and is strongly related to the $L/L_{Edd}$ ratio, as shown by
disk-corona models, and strengthed by intrinsic reddening.}
\item{Eigenvector~2 (18\% of variance) correlates with optical/IR
colors and optical type, and, as shown by the optical/IR color
modeling, is related to the relative strength of the host galaxy to
the observed/reddened AGN.}
\item{Eigenvector~3 (12\% of variance) correlates with reddening
indicators obtained from X-rays and narrow emission lines, indicating
a common absorber that affects both the X-rays and narrow emission
lines and probably lies in a moderately inclined host galaxy.}
\item{Eigenvector~4 (8\% of variance) correlates with the degree of
polarization and broad H$\alpha$/H$\beta$ ratio implying the presence
of dust (different from the host galaxy absorber in EV3) that scatters
both the BLR and the continuum emission and reddens the broad lines.}

\end{itemize}

PCA analysis of the SED and/or emission line properties of unobscured
(BG92) and moderately obscured AGN (Marziani \etal\ 2001, red 2MASS
sample presented here) points to the $L/L_{Edd}$ ratio as the most
important parameter in determining the SED and emission line
properties in AGN.  Since the red 2MASS sample has moderate
obscuration, EV1 is also slightly effected by reddening, as is the EV1
found by Marziani \etal\ (2001) in their heterogeneous sample of
moderately obscured AGN.  Successive eigenvectors pick up components
depending on the sample chosen: unobscured AGN samples (BG92) probe
the properties of the bright central engine (BG92 EV2 is related to
accretion rate - see Boroson 2002).  Moderately obscured samples, such
as the red 2MASS AGN, pick up weaker components, like the host galaxy
emission, which become important as the reddening obscures the AGN
emission. Spectral PCA of the moderately obscured SDSS QSOs (Yip
\etal\ 2004) also finds its EV2 to be related to the host galaxy.
Successive eigenvectors in the red 2MASS sample are related to dust
obscuration, separated into circumnuclear and host galaxy
absorption. Our analysis is consistent with the orientation-dependent
unification scheme for AGN, where a range of inclination dependent
obscuration and host galaxy contribution (as suggested by Haas \etal\
2008) and most importantly $L/L_{Edd}$ ratios (as suggested by our
PCA) can explain the properties of an AGN.

\acknowledgements

We wish to thank the referee, Gordon Richards, for comments that helped
improve the paper and Martin Gaskell for useful discussions. BJW and
JK gratefully acknowledge the financial support of NASA Chandra
grants: GO1-2112A, GO3-4138A and NASA XMM-Newton grants: NNG04GD27G,
NNG05GM24G, which supported various aspects of this work.  We also
gratefully acknowledge the financial support of grants: NAS8-39073,
GO-09161.05-A (HST). PSS acknowledges support from NASA/JPL contract
1256424. This publication makes use of data products from the Two
Micron All Sky Survey, which is a joint project of the University of
Massachusetts and the Infrared Processing and Analysis
Center/California Institute of Technology, funded by the National
Aeronautics and Space Administration and the National Science
Foundation.  We wish to thank F. Murtagh for making his PCA program
available on the web.

\newpage


\clearpage

\clearpage

\clearpage

\clearpage

\clearpage

\clearpage

\clearpage

\clearpage
\begin{figure}
\plotone{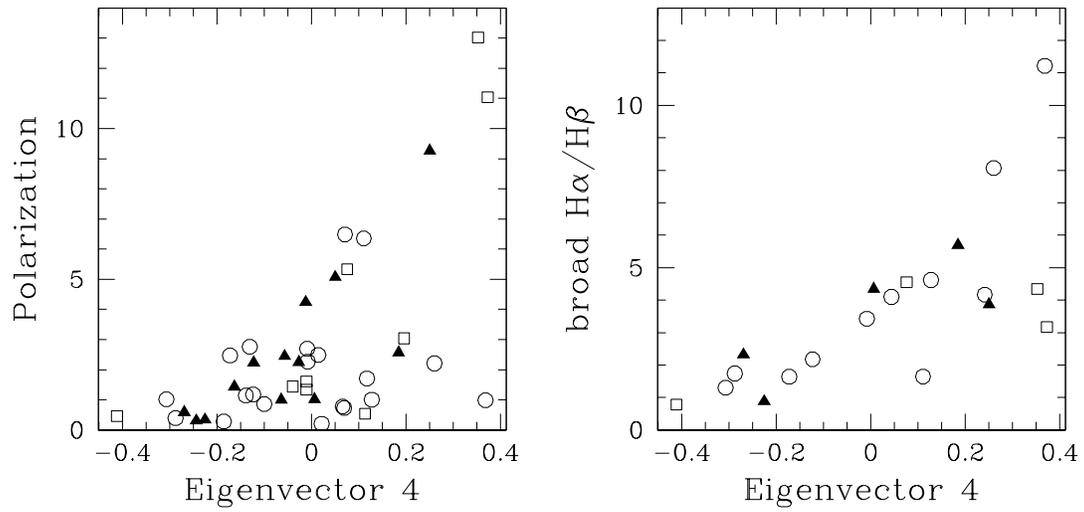}
\caption{Correlations between Eigenvector~4 and the degree of optical
polarization and broad H$\alpha$/H$\beta$. Symbols indicate X-ray
S/N, as in Fig.~\ref{fig:Fx/FBvsFx/Fnn}.}
\label{fig:EV4}
\end{figure}


\clearpage
\begin{deluxetable}{lrrrr}
\tablenum{1}
\tablecaption{Correlation of eigenvectors with SED and emission line parameters.}
\tablewidth{0pt}
\tablehead{
\colhead{Variable} &
\colhead{EV1} &
\colhead{EV2} &
\colhead{EV3} &
\colhead{EV4} \\
\colhead{} &
\colhead{32.9\%} &
\colhead{17.8\%} &
\colhead{11.6\%} &
\colhead{8.0\%}
}
\startdata 
 Type                            &   0.0709 &{\underline {\bf $-$0.3933}} &  0.0914 &$-$0.3100 \\ 
 Polarization                    & $-$0.0358 &$-$0.1634 &$-$0.1641 & {\underline {\bf 0.4851}} \\
 $B-R$                           &   0.1543 &{\underline {\bf $-$0.4196}} &$-$0.0003 &  0.0992 \\ 
 $J-K$                           & $-$0.0563 &{\underline {\bf $-$0.3591}} &$-$0.3130 &  0.0368 \\
 $B-K$                           &   0.0678 &{\underline {\bf $-$0.4521}} &$-$0.2056 &  0.0547 \\
 $N_H$                           & $-$0.0057 &$-$0.2166 &{\underline {\bf  0.3696}} &$-$0.0887 \\
 $\alpha_{opt}$                  & $-$0.2108 &  0.3248 &  0.0724 &  0.1596 \\
 $\Gamma_X$                      &   0.0468 &  0.1529 &{\underline {\bf $-$0.3814}} &$-$0.2999 \\
 z                               & $-$0.0914 &$-$0.0933 &$-$0.3250 & {\underline {\bf 0.4314}} \\
 $\log F_{1keV}/F_B$             &{\underline {\bf 0.4035}} &$-$0.0302 &$-$0.0100 &  0.0291 \\
 $\log F_{1keV}/F_R$             &{\underline {\bf 0.3999}} &  0.0454 &  0.0172 &  0.0147 \\
 $\log F_{1keV}/F_I$             &{\underline {\bf 0.3894}} &  0.0611 &  0.0415 &  0.0800 \\
 $\log F_{1keV}/F_J$             &{\underline {\bf 0.3861}} &  0.1022 &  0.0230 &  0.0757 \\
 $\log F_{1keV}/F_K$             &{\underline {\bf 0.3742}} &  0.1686 &  0.0862 &  0.0822 \\
 Hardness ratio                  & $-$0.0520 &$-$0.2803 &{\underline {\bf  0.4958}} &$-$0.0517 \\
 H$\alpha^{NLR}$/H$\beta^{NLR}$  & $-$0.1120 &  0.0069 &{\underline {\bf  0.3347}} &  0.1694 \\
 H$\alpha^{BLR}$/H$\beta^{BLR}$  & $-$0.0612 &  0.0165 &  0.2509 &{\underline {\bf  0.5241}} \\
 $[OIII]/F_{2-10keV}$		 &{\underline {\bf $-$0.3531}} &$-$0.0104 &$-$0.0227 &$-$0.1378 \\
\enddata

\tablecomments{Parameters that dominate an eigenvector are denoted in
  bold font and underlined.}
\end{deluxetable}


\begin{table}
\label{tab:PCA} 
\end{table}


\clearpage
\begin{figure}
\plotfiddle{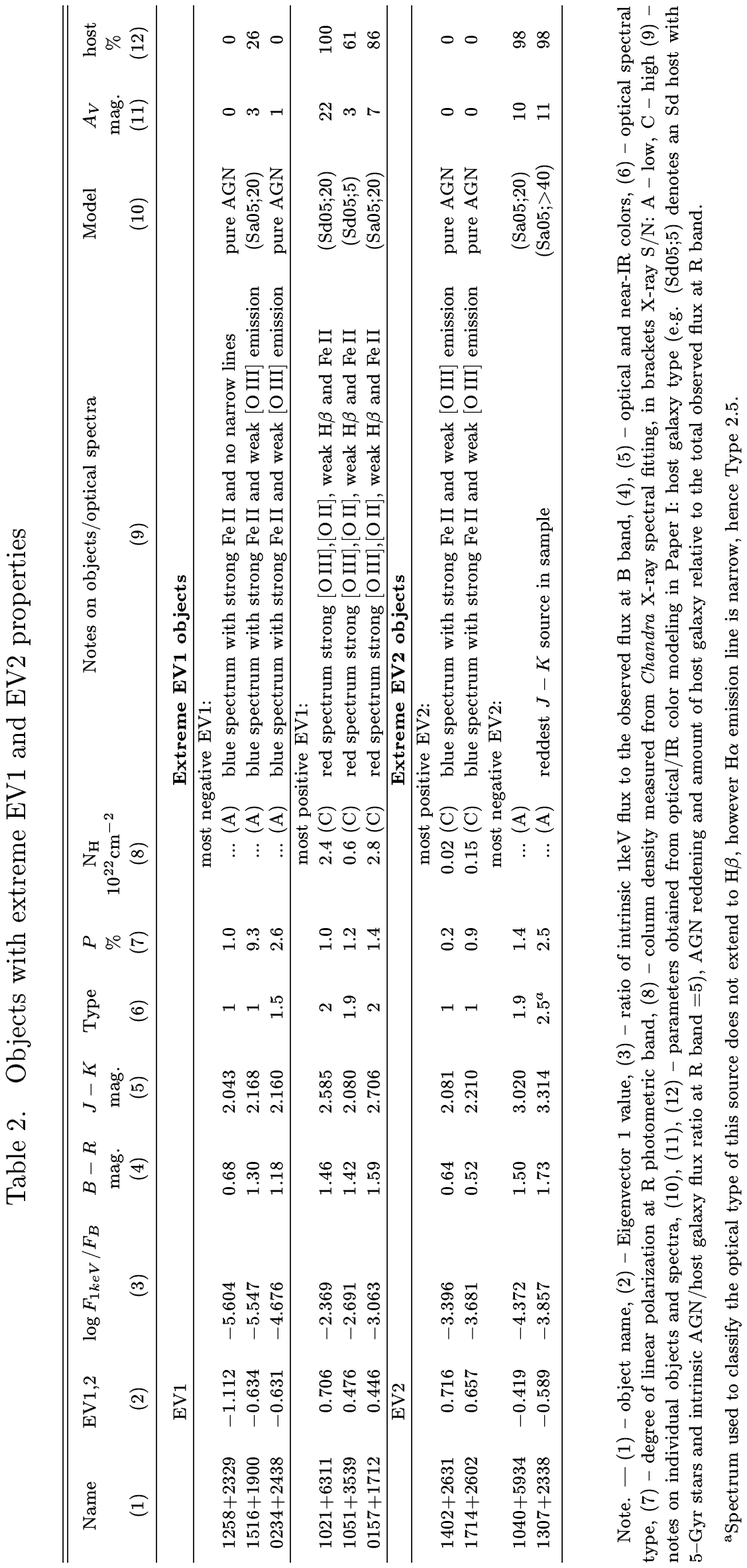}{6in}{0.}{90}{90}{-240}{-250} 
\label{tab:extremeEV} 
\end{figure}


\begin{references}

\reference{} Alonso-Herrero, A., Ward, M. J., Kotilainen, J. K. 1997,
MNRAS, 288, 977 



\reference{} Boller, Th., Brandt, W. N., \& Fink, H. 1996, A\&A, 305, 53

\reference{} Boroson, T. A., \& Green, R. F. 1992, ApJS, 80, 109 (BG92)

\reference{} Boroson, T. A. 2002, ApJ, 565, 78

\reference{} Brandt, N., \& Boller, Th. 1998, AN, 319, 163 

\reference{} Buzzoni, A. 2005, MNRAS, 361, 725

\reference{} Corbin, M. R. 1993 ApJ, 403, L9

\reference{} Cutri, R., Nelson, B. O., Francis, P. J. \& Smith, P. S.
2002, ``AGN Surveys'', Edited by R. F. Green, E. Y. Khachikian, and
D. B. Sanders, Proceedings of IAU Colloquium 184, ASP Conference
Proceedings, Vol. 284, p. 127
	
\reference{} Deo, R. P., Crenshaw, D. M., Kraemer, S. B., Dietrich,
M., Elitzur, M., Teplitz, H., Turner, T. J. 2007, ApJ, 671, 124

\reference{} Elvis M., Wilkes, B. J., McDowell, J. C., Green, R. F.,
Bechtold, J., Willner, S. P., Oey, M. S., Polomski, E., \& Cutri, R. 1994,
ApJS, 95, 1

\reference{} Francis, P.J. \& Wills, B. J., 1999, ``Quasars and
Cosmology'', eds. G. Ferland \& J. Baldwin, ASP Conference Series,
Vol. 162, 363
	
\reference{} Francis, P. J., Hewett, P. C., Foltz, C. B., Chaffee,
F. H., 1992, ApJ, 398, 476

\reference{} Haas, M., Willner, P. S., Heymann, F., Ashby M. L. N.,
Fazio, G. G., Wilkes B. J., Chini, R., Siebenmorgen, R., 2008,
astro--ph/0807.3966 

\reference{} Jackson, N., \& Browne, I. W. A. 1990, Nature, 343, 43

\reference{} Krolik, J. H., McKee, C. F., Tarter, C. B. 1981, ApJ, 249, 
422  

\reference{} Kuraszkiewicz Wilkes, B. J., Brandt, W. N., \&
            Vestergaard, M., 2000, ApJ, 542, 631

\reference{} Kuraszkiewicz, J. K., Wilkes, B. J., Hooper, E. J.,
McLeod, K. K., Wood, K., Bjorkman, J., Delain, K. M., Hughes, D. H.,
Elvis, M. S., Impey, C. D., Lonsdale, C. J., Malkan, M. A.,
McDowell, J. C., \& Whitney, B. 2003, ApJ, 590, 128

\reference{} Kuraszkiewicz, J. K., Wilkes, B., Schmidt, G., Ghosh, H.,
Smith, P. S., Cutri, R., Hines, D., Huff, E. M., McDowell, J. C.,
Nelson, B., 2008, ApJ submitted

\reference{} Laor, A., Fiore, F., Elvis, M., Wilkes, B. J., \& McDowell,
J. C. 1994, ApJ, 435, 611 

\reference{} Laor, A., Fiore, F., Elvis, M., Wilkes, B. J., \& McDowell,
J. C. 1997, ApJ, 477, 93 

\reference{} Marble, A. R., Hines, D. C., Schmidt, G. D.,
Smith, P. S., Surace, J. A., Armus, L., Cutri, R. M., Nelson, B. 2003
ApJ, 590, 707

\reference{} Marziani, P., Sulentic, J. W., Zwitter, T., Dultzin-Hacyan, D.,
Calvani, M., 2001, ApJ, 558, 553

\reference{} Mulchaey, J. S., Koratkar, A., Ward, M. J.,
Wilson, A. S., Whittle, M., Antonucci, R. R. J., Kinney, A. L.,
Hurt, T. 1994, ApJ, 436, 586

\reference{} Murray, N. \& Chiang, J. 1997, ApJ, 494, 125 

\reference{} Murtagh, F. \& Heck, A. 1987, ``Multivariate Data
Analysis'' Springer-Verlag Berlin Heidelberg 1987, ISBN 90-277-2425-3 


\reference{} Polletta, M., Weedman, D., Hönig, S., Lonsdale, C. J.,
Smith, H. E., Houck, J. 2008, ApJ, 675, 960

\reference{} Pounds, K. A., Done, C., \& Osborne J. 1995, MNRAS, 277, L5

\reference{} Proga, D. 2005, ApJ, 630, L9

\reference{} di Serego Alighieri, S., Cimatti, A., Fosbury, R. A. E.,
\& Hes, R. 1997, A\&A, 328, 510

\reference{} Shang, Z., Wills, B. J., Robinson, E. L., Wills, D.,
Laor, A., Xie, B.; Yuan, J, 2003, ApJ, 586, 52

\reference{} Smith, P. S., Schmidt, G. D., Hines, D. C., Cutri, R. M.,
Nelson, B. O. 2002, ApJ, 569, 23

\reference{} Smith, P. S., Schmidt, G. D., Hines, D. C.,
Foltz, C. B. 2003, ApJ, 593, 676

\reference{} Sulentic  J. W., Zwitter, T., Marziani, P.,
Dultzin-Hacyan, D. 2000, ApJ, 536, L5

\reference{} Tadhunter, C. N., Morganti, R., Robinson, A., Dickson, R.,
Villar-Martin, M., \& Fosbury, R. A. 1998, MNRAS, 298, 1035

\reference{} Turner, T.J., George, I.M., Nandra, K. \& Mushotzky,
R.F. 1997, ApJ, 113, 23

\reference{} Wang, T.-G., Brinkmann, W., \& Bergeron, J. 1996, A\&A, 309, 81

\reference{} Wang, J., Wei, J. Y., \& He, X. T. 2006, ApJ, 638, 106

\reference{} Wilkes, B. J., Schmidt, G. D., Cutri, R. M., Ghosh, H.,
Hines, D. C., Nelson, B., Smith, P. S. 2002, ApJ, 564, L65

\reference{} Wills, B.J.,  Shang, Z., \& Yuan, J. M., 2000, NewAR, 44, 511

\reference{} Wills, B.J., Laor, A., Brotherton, M. S., Wills, D.,
Wilkes, B. J., Ferland, G. J., Shang, Z., 1999, ApJ, 512, L53 

\reference{} Witt, H. J. Czerny, B., \& \.Zycki, P. T. 1997, MNRAS, 286, 848

\reference{} Yip, C. W., Connolly, A. J., Vanden Berk, D. E., Ma, Z.,
Frieman, J. A., SubbaRao, M., Szalay, A. S., Richards, G. T.,
Hall, P. B., Schneider, D. P., 2004, AJ, 128, 2603



\end{references}
\end{document}